# The 3D genome shapes the regulatory code of developmental genes

Julien Mozziconacci[1,2*], Mélody Merle[1] & Annick Lesne[1,3,*]

[1]*Sorbonne Université, CNRS, Laboratoire de Physique Théorique de la Matière Condensée, LPTMC, F-75252, Paris, France*

[2]*Muséum National d'Histoire Naturelle, Structure et Instabilité des Génomes, UMR7196, 75231, Paris Cedex 5*

[3]*Institut de Génétique Moléculaire de Montpellier, University of Montpellier, CNRS, Montpellier, France*

** Corresponding authors: mozziconacci@lptmc.jussieu.fr, lesne@lptmc.jussieu.fr*

**Summary** *(199 mots)*

We revisit the notion of *gene regulatory code* in embryonic development in the light of recent findings about genome spatial organisation. By analogy with the genetic code, we posit that the concept of *code* can only be used if the corresponding *adaptor* can clearly be identified. An adaptor is here defined as an intermediary physical entity mediating the correspondence between *codewords* and *objects* in a gratuitous and evolvable way. In the context of the gene regulatory code, the encoded objects are the gene expression levels, while the concentrations of specific transcription factors in the cell nucleus provide the codewords. The notion of code is meaningful in the absence of direct physicochemical relationships between the objects and the codewords, when the mediation by an adaptor is required. We propose that a plausible adaptor for this code is the gene domain, that is, the genome segment delimited by topological insulators and comprising the gene and its enhancer regulatory sequences. We review recent evidences, based on genome-wide chromosome conformation capture experiments, showing that preferential contact domains found in metazoan genomes are the physical traces of gene domains. Accordingly, genome 3D folding plays a direct role in shaping the developmental gene regulatory code.

**Abbreviations**: TAD: topologically associating domain; TF, transcription factor; tRNA, transfer RNA;

**Highlights:**

1) A code is defined by a set of adaptors, in the sense introduced by Crick for tRNAs

2) The gene regulatory code relates TF concentrations to gene expression levels

3) The adaptor of the gene regulatory code is a 3D physical unit: the gene domain

4) An example is the regulation of eve by gap proteins in the Drosophila embryo

**Graphical abstract**

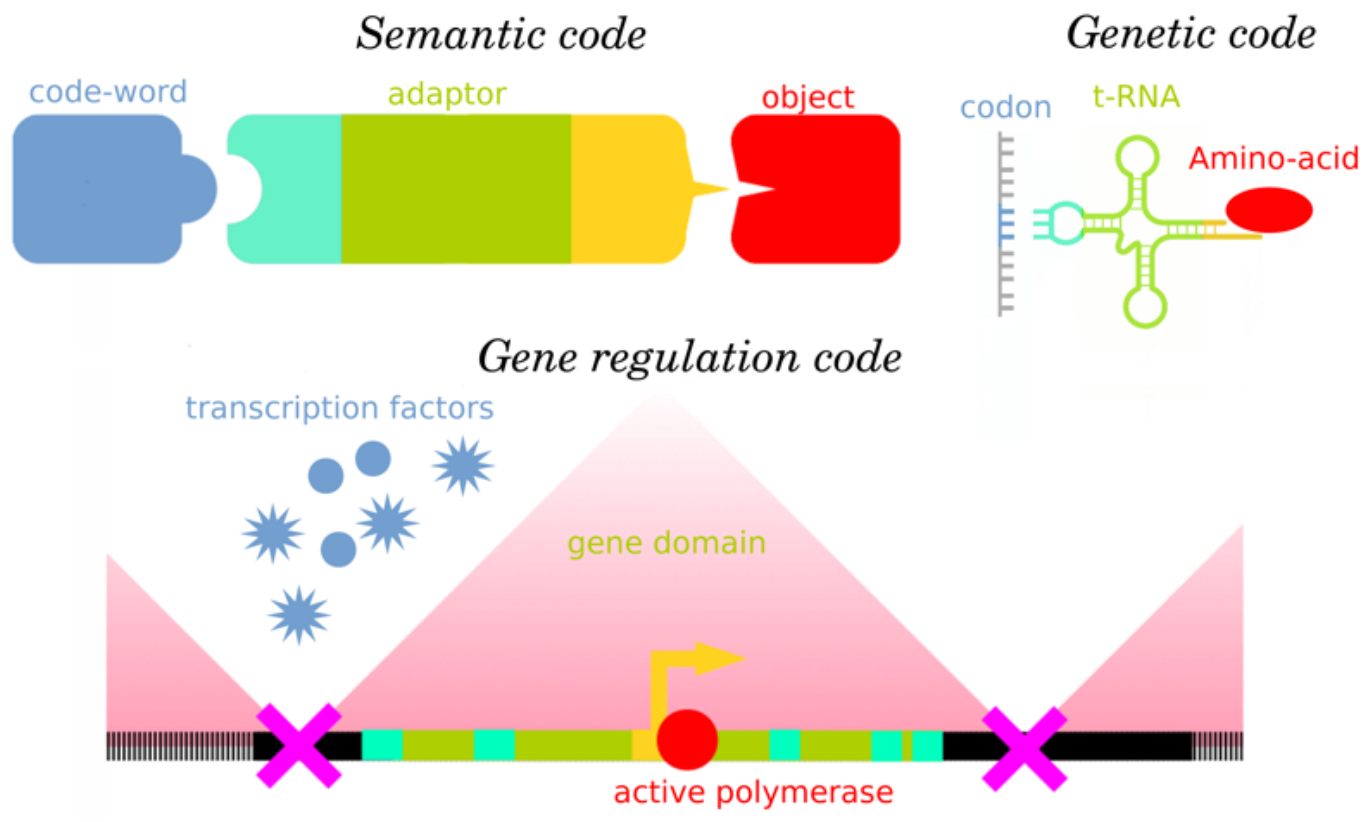

## 1. Introduction

The development of a multicellular organism from a unique cell, the fertilised egg, centrally relies on cell differentiation. In this process, local differences in gene expression lead to the formation of distinct cell types in different places in the embryo. Each gene responds to the local concentrations of transcription factors (TFs) in a way that determines its own transcriptional output. The initial symmetry breaking that leads to different TF concentrations in different cells of the embryo varies from one organism to another [1]. In Drosophila, for instance, the diffusion of maternal factors from one pole of the embryo to the other generates gradients that account for a first spatial heterogeneity in TF concentrations, along the anterior-posterior axis. Once this first patterning is formed, a cascade of genes is activated resulting in robust spatial distribution of gene expression. This distribution, which is the basis of the future body plan, thus critically depends on regulatory dependencies between genes. A complex combinatorics relates TF concentrations and the target gene expression [2-4]. Several studies have addressed the potential existence of a *gene regulatory code*, that would regulate gene expression at the transcriptional level. The very term of gene regulatory code has been introduced in studies looking for an association between regulatory DNA sequences and spatial gene expression patterns [5, 6]. The term cis-regulatory code has more recently been used to sum up the influence on gene expression of regulatory sequences and their linear organisation along the genome [7]. We will here critically re-evaluate these notions and propose that a carefully defined notion of code is instrumental for interpreting most recent experimental data.

Our starting point is the definition of a *code* that will be used in the present text. Different meanings of this word are encountered in science, from the secret codes in cryptography, the source codes in computer science, to the neural codes and the genetic code. The latter is the emblematic example of a *semantic code*, in a biological context (Fig. 1A). The definition of a semantic code relies on three ingredients, namely codewords, objects, and adaptors (Fig. 1B):

- codewords are inputs to be interpreted;
- a single object is associated to each codeword;
- adaptors are physical entities that implement the association of each codeword with a unique object, which can be seen as 'the object encoded by the codeword'.

Semantic codes go beyond a catchword, as their ability to evolve and their gratuity, in the sense introduced by Monod [8], provide evolutionary and mechanistic insights. To clarify the aforementioned relationships between TF concentrations, enhancers and gene expression, we will revisit the notion of gene regulatory code, *taken with a semantic meaning.* In this context, the codewords would be TF concentrations while the objects would be the expression status of a gene, with the idea that a particular set of TF concentrations elicits a defined transcriptional output. The missing link required to determine a gene regulatory code is the identification of the adaptor. We will base our analysis on recent experimental quantifications of gene expression during Drosophila embryogenesis and insights coming from the detailed study of 3D genome organisation in the surroundings of genes.

**2. A semantic code is characterised by its adaptors**

During the early years of the discovery of the genetic code, Crick [9] proposed that some molecular entities should exist that relate codons to amino-acids. They are known today to be tRNAs. Crick originally called them *adaptors* (and actually argued that they could be made of RNA). This term of adaptor is particularly meaningful, as it emphasises the mediating role of this set of molecules: each tRNA binds a codon and the corresponding amino acid at two different sites (Fig. 1A). The emblematic example of the genetic code actually meets the definition of a semantic code, as a *conventional* correspondence between a set of objects and a set of codewords, without any particular correspondence being imposed by the laws of nature [10]. In a biological context, a correspondence between two molecular entities thus deserves to be termed a code when mediated by molecular adaptors that are capable of recognising separately each of these entities (Fig. 1B). By its very existence, the adaptor circumvents the need of direct physical, chemical or stereo-chemical interactions between the two entities. Which one of them is the codeword (the other being the encoded object) follows from the functional status of the correspondence. For instance, in the case of the genetic code, the codewords encoding the amino acids at the genomic level are the codons.

Identifying the adaptors demonstrates that speaking of a code is not a mere metaphor: it offers mechanistic insights on the evolutionary origin of the code. The code is conventional in the sense that the relationship between codewords and objects does not result from some direct and unavoidable physicochemical interaction, but from the mediation of an adaptor. In principle, any pair composed of a codeword and an object could be related once the proper adaptor has been produced in the course of molecular coevolution. The arbitrariness of a code implies its ability to evolve. Indeed, the mapping relating codewords and encoded objects can be changed by a change in the adaptor. This offers a way to devise synthetic variants, as done for the genetic code [11].

Given this operational definition, it appears that the word *code* is often misleadingly used in the scientific literature. In particular, several so-called 'codes' have been introduced in a biological context at the molecular level, for instance sequence codes [12] or nucleosome codes [13]. Sequence codes are based on the affinity between base pairs A/T and G/C, i.e. the physically prescribed Watson-Crick pairing rule, and not on a gratuitous association, as required for a semantic code. Similarly, nucleosome codes are based on the physicochemical affinity between DNA sequence and the histone core, and do not involve any adaptor. In these cases, speaking of a code is only metaphoric, based on the plain meaning of the word in common speech. In short, calling some association a code is pointless when a modification in the laws of physics and chemistry would be required to change this association.

## 3. A domain including the gene promoter and its enhancers is a plausible adaptor of the gene regulatory code

fd by the above caveats about the relevance (or irrelevance) of naming a relationship a code, we will examine published facts on transcriptional regulation, in the context of development. The discussion in the previous section provides a critical guideline: the existence of a gene regulatory code would be assessed by identifying its adaptor.

Pioneering studies investigating the early development of sea urchin have led to the discovery of genomic sequences, lying beside a developmental gene, which are necessary and sufficient to determine the transcription of this gene in response to the proper external cues. These sequences spanned a 2kb region, which was then referred to as the gene regulatory domain in the seminal publication [14]. Many such gene regulatory domains were later identified in different organisms. Their genetic dissection led to the identification of specific short sequences of ~100 bp that are rich in TF binding sites and essential for gene expression. These sequences were shown to coincide with sequences today known as *enhancers*, originally discovered in a different context [15]. Enhancers are now recognised as the basic ingredient of the developmental gene regulation in metazoans (see [16] for a review). The current understanding of their action involves the formation of an enhancer-promoter loop [17, 18]. While the simplest scheme involves only one TF, clusters of TF binding sites are usually found at an enhancer [7]. Some TFs can be activators, i.e. their binding to the enhancer is a prerequisite to transcription initiation, whereas other TFs are inhibitors, i.e. their binding to the enhancer prevents transcription initiation [7]. TF can even have dual functions [19]. Different combinations of TF concentrations are then associated with different transcription outputs. However, single-enhancer action is not sufficient to explain gene patterning during development. Most often genes are under the control of several enhancers [20]. In parallel, it has been observed in Drosophila embryo that different combinations of TF concentrations can be associated with the same level of expression a developmental gene [21, 22].

We thus propose that a plausible adaptor of a gene regulatory code is provided by an extension of the primary gene regulatory domain notion, that we will simply name the *gene domain* (Figs. 1-3). In contrast to the primary notion, the gene domain contains the gene enhancers, but also its promoter, as well as the linker sequences between them, overall forming a single composite unit able to both read TF concentrations and trigger gene transcription. The parallel with the general notion of adaptor is underlined by using the same colours in Figs. 1-3.

Additional players, the insulators, sets the limits of the gene domain along the genome sequence (Fig. 3). An insulator has been operationally defined as a sequence located between a gene and an enhancer, which prevents the spurious action of the enhancer onto this gene [23]. Initially discovered in Drosophila, insulators have then been characterised in many, if not all, metazoans, including human [24].

**4. Gene domains can be physically identified as preferential contact domains**

Additional insights into the regulation of developmental genes came in the last years from chromosome conformational capture experiments (3C). This ensemble of techniques has been devised to investigate the three-dimensional (3D) folding of the genome in living cells [25]. They consist in a chemical crosslinking of protein-mediated contacts between pairs of genomic loci, followed by a molecular dissection of the crosslinks and a sequencing-based identification of the genomic loci involved. They accordingly provide contact frequencies between genome regions, reflecting the in-vivo 3D organisation of these regions in the nucleus [17, 26]. In particular, they have shown that, most often, insulators prevent physical contacts between the genomic loci they insulate [23, 27]. 3D genome folding reflects the proper physical communication between the enhancers and the gene promoter, while it prevents enhancer-promoter contacts across the insulators [28]. An important experiment in this context has been to use hundreds of random insertions of an enhancer at different positions along the genome and to measure gene expression levels in the vicinity of the insertion. The results showed that the activity of the enhancer is homogeneous within wide domains separated by insulators [29]. Accordingly, as outlined on Fig. 3A, the gene domain can consistently be defined as the genome domain, located between two insulators, that contains the gene and its enhancer(s).

Preferential and prevented contacts can be mapped along the genome using the genome-wide derivative of 3C (Hi-C). The first preferential contact domains identified with Hi-C were called Topologically Associating Domains (TADs) [30-32]. These domains are on average 1Mb long (in mammals) and usually contain several gene domains. With the increase of the resolution of Hi-C maps, sub-TADs and even smaller micro-TADs have been identified within TADs [27,33,34]. These smaller contact domains would correspond to the gene domains as defined here. An ever-increasing number of functional evidences indeed suggests an important role of TAD boundaries in regulating genes in space and time during development.

**5. Assessing the effect of preferential contact domains on gene transcription**

In order to investigate the general role of TADs and sub/micro-TADs on gene expression, two different strategies have been employed. In global strategies the domains are disrupted genome-wide through the perturbation of one of the players that drive domain formation. In local strategies, a specific genomic element, usually containing a domain boundary, is modified.

In mammals, two important players have been identified for domain formation. The first is CTCF, which binds at domain borders. The second is cohesin, which induces the formation of a contact domain between two CTCF sites. Several studies investigated the effect of cohesin and/or CTCF depletion on gene expression at the genome-wide level [35-37]. Their results quantitatively depend on the cell type used and the strategy for protein depletion but all these studies pointed out an effect of the contact domains on gene regulation. While this effect is small or non-existent for housekeeping genes, it is stronger for cell-specific genes, suggesting that gene domains as defined by contact domains are more relevant to describe developmental gene regulation [38].

The prevalent role of contact domains in regulating genes during development has been evidenced on specific examples identified using genomic variants known to induce pathogenic phenotypes [39-41]. Disruption of a contact domain boundary may induce ectopic enhancer-promoter interactions causing developmental anomalies in mouse [40,42]. As a paradigmatic example, deletions, inversions, or duplications of genomic segments containing a contact-domain boundary induce the mis-regulation of genes at the Epha4 locus [43]. A similar effect has been demonstrated at the level of a single developmental gene, where changes in 3D chromatin conformation may restrict the enhancer activity and result in gene mis-expression [44]. In a very different context, the hypermethylation of a specific insulator has been shown to prevent CTCF function, leading again to gene mis-expression and cancer [45]. This link between domain alteration and cancer has been further exploited to provide pan-cancer candidate genes [46]. More and more developmental loci are now under scrutiny and it is becoming increasingly clear that while the underlying mechanisms still need to be investigated [47]. contact domains have an important role in regulating genes at the right time and in the right place during development. On the other hand, the functional impact of contact domains appears less clear when considering gene regulation at a genome-wide level in cell cultures or tissues (see other reviews on that matter in this special issue [48, 49].

In the gene domain model, genes are supposed to be completely insulated from others. TADs (on average 1Mb long in mammals) usually contain several gene domains, appearing in Hi-C maps as sub-TADs. Exploring the detailed relationship between TADs and gene domains thus offers a way towards a better understanding of gene co-expression. In this line, it has been proposed that TADs form functional units, each composed of co-regulated genes [50, 51].

### 6. Gene domains allow for an epigenetic control of the developmental code

The association between codewords and transcriptional responses is achieved by the gene domain with some epigenetic plasticity. Indeed, this association can be modified, following either a change in the borders of the gene domain that would add enhancers in (or remove from) the domain [52], or a change in the sequence of an enhancer affecting its affinity for specific TFs. A dramatic example of this latter possibility is given by the case of the ZRS enhancer of the *Sonic hedgehog (Shh)* gene, where a synthetic restauration of a single ETS1 binding site can re-establish a normal phenotype of otherwise serpentinised mouse (i.e. a mouse with truncated limbs) [53]. An example of the first possibility is observed in the case of HoxD gene cluster in mice, which lies between two TADs [50,54]. A shift in the contacts established by Hox genes is observed at different positions in the embryo and different times during development. The gene domains are accordingly modified possibly impacting the expression level of the target genes. Similar changes can be observed by engineering appropriate deletions in the gene domain boundary [55]. The increasing resolution of Hi-C maps should help in the future to precisely delineate gene domains and quantitatively confirm these mechanisms.

### 7. Gene domains allow for gene regulation by multiple enhancers

The notion of gene domain as an adaptor encoding the gene transcriptional regulation goes beyond the current picture of enhancer-based regulation, presented on Fig. 2. Indeed, in eukaryotes, it appears that most developmental genes are under the control of several enhancers [7], which confers pleiotropy and multi-functionality to the genes [40]. In contrast, in prokaryote unicellular organisms, transcriptional regulation relies on regulatory sequences located close to the gene transcription start site [56]; genomes are compact, with no need of a code nor of a gene domain to integrate multiple cues.

In multicellular organisms, which have longer genomes, gene domains embed several enhancers and integrate of their regulatory activities. Fig. 3B illustrates the combinatorics of molecular cues and associated responses in the simple case of two TFs, two enhancers and an all-or-none control. In the case presented there, the activity of the two enhancers is coupled with an exclusive OR logical function, so that the gene is active if one and only one of the two TFs is present. In real situations, the response is gradual. The corresponding case of a continuous control is sketched on Fig. 3C. We intro-duce a 2D heat map where each axis represents a TF concentration and the colour scale represents the gene expression level (typically, the normalised mRNA level), and propose to call such a plot the *expression phase diagram* of a gene. This diagram is expected to display well-defined regions in which the target gene is significantly expressed (active phases) and regions where the gene remains silent (inactive phases), each of these regions are defined by a set of codewords. The expression phase diagram is thus the most relevant piece of information to explicitly describe the gene regulatory code.

The way enhancers interact with each other is sometimes unclear. Any kind of logical or analytical function could in principle be envisioned, and efforts have been made both experimentally [57] and theoretically [58] to determine whether enhancers act additively, competitively or in another way. As a special case, shadow enhancers have been proposed to act as redundant enhancers in order to add robustness to the decoding process [59].

**8. Codewords can be used to read positional information within the embryo.**

In the discussion above, we posit that codewords for the gene regulation code are concentrations of few specific TFs, since this fits in the general scheme of a semantic code. This idea that local TF concentrations regulate gene expression during development goes back to the concept of *positional information* introduced in 1968 by Lewis Wolpert. He proposed that some local feature (in his words, a positional value) could underlie the spatio-temporal pattern of cell differentiation in the Drosophila embryo [60-62]. Since then, the concept of positional information has been extensively studied and its application to complex gene expression patterns was put forward early on [63]. Nevertheless, the quantitative analysis of positional information only began fifteen years later. Using fine in-vivo biophysical measures, it was shown that local concentrations can be defined sharply enough in the embryo to play the role of codewords (in this case simple codewords since only one TF was considered) [64]. A remarkable finding was that the accuracy with which the concentration profiles are established is remarkably good, with less noise than previously expected.

The combination of the two notions of positional information and expression phase diagram has practical implications when considering spatial gene regulation within the developing embryo. Fig. 4 sketches a gene regulation scenario for a simplified embryo, in which the target gene is controlled by two TFs according to the truth table in Fig. 3B. If one assumes the spatial distributions of TF concentrations depicted in Fig. 4A, the red gene regulated by these two TFs is expressed in three stripes (Fig. 4B). In stripes 1 and 2, the activation results from the use of the same codeword whereas in stripe 3 it results from the use of a different codeword as sketched on the expression phase diagram reproduced in Fig. 4C.

**9. A real-world example: the gene regulatory code of *Eve* in the early patterning of Drosophila embryos.**

In Drosophila early development, the nucleus of the fertilised egg undergoes several rounds of division forming a syncytium, i.e. a large multinucleate cell in which about 6000 nuclei occupy distinct positions in space. The first layer of differentiation is based on the gradient of maternal factor concentrations, which are already asymmetrically distributed in the egg. They impact the transcription of four elementary genes, *Krüppel, Hunchback, Giant* and *Knirps,* globally known as *gap* genes. This in turn produces a heterogeneous distribution of the associated proteins. These proteins act as TFs for other developmental genes, known as *pair-rule* genes, which exhibit a stripe-patterned expression [65]. The *even-skipped* (*eve)* gene belongs to the family of pair-rule genes and displays a seven-striped pattern (Fig. 5). While this emblematic gene has been so far the most studied, other pair-rule genes behave in a similar way [66]. The *eve* gene domain, which contains five enhancers [67], is able as a whole to properly respond to the different combinations of gap-protein concentrations encountered in the embryo. Two recent studies used the Drosophila melanogaster embryo to show that the spatial expression (protein concentration) of the *eve* gene can be quantitatively predicted from the concentration patterns of the four gap TFs Krüppel, Hunchback, Giant and Knirps [22,68]. Being robustly defined and associated with a unique eve expression response, the set of these four TFs concentrations could thus provide in principle the input of the gene regulatory code involved in *eve* early patterning.

To illustrate further this assertion, we performed a complementary analysis of other recent experimental data (Figs. 5 and 6). We used the quantitative measurements of mRNA levels in each nucleus of the Drosophila embryo at stage 14 provided by the Berkeley 3D gene expression atlas [21,69]. For our purpose, the spatial distribution of mRNAs in the embryo offers a way to sample at different time points the various combinations of TF concentrations as well as the associated expression of *eve.* Indeed, when both the protein and mRNA levels of a developmental gene were available, we have checked that they display a high correlation of 0.89, that could even reach 0.95 when optimally tuning the time delay between the protein and mRNA measurements. We thus extracted the (normalized) mRNA concentrations of Krüppel, Hunchback, Giant and Knirps (Fig. 5, top) in the nuclei located in each of the seven stripes of *eve* expression (Fig. 5, bottom). The result is presented in the form of a box-plot for each stripe (Fig. 5, middle). We checked its robustness with respect to the expression threshold used to delineate *eve* stripes. The plot shows quantitatively that each stripe is associated with a well-defined codeword. A more precise test of the gene expression code scenario could be done in a near future using space- and time-resolved *eve* transcription data [70] instead of averaged concentrations of the gene products (mRNA). Joint spatial imaging of genome folding and gene ex-pression at the single-cell level will be instrumental to quantitatively delineate the cell-to-cell variability of the relationship between 3D domains and transcriptional activity [71-73]. This should ultimately explain how intrinsic stochasticity of molecular processes can be canalized into a robust average gene expression at the organism level.

Dedicated experiments have shown that one of the enhancers is predominantly involved in the formation of each stripe. However, these experiments also revealed that the additive modularity and one-to-one correspondence between enhancers and stripes is not exact. For instance, Lim *et al*. removed stripe 1 using a CRISPR/Cas9-mediated deletion on the corresponding enhancer, but this modification also induced a displacement of stripe 2 [74]. Kim *et al*. constructed several *eve* loci variants containing only the enhancers for stripe 2 and the pair of stripes 3 and 7. They showed that the order of the enhancers in the fusion construct significantly affects the expression levels [75]. The gene domain is thus more than the sum of its enhancers and we need to consider its coordinating action to account for the target gene regulation.

An integrated view of the gene regulatory code associated with the gene domain of *eve* is presented on Fig. 6. We again used the normalised mRNA concentrations of Krüppel, Hunchback, Giant, Knirps and Eve, Fig. 5, to draw a regulatory phase diagram similar in principle to the one shown in Fig. 4C, now using real data. In this case, the TF concentration space is now four-dimensional, which requires the use of a projection method, known as multidimensional scaling [76], for the representation in the two dimensions of a paper sheet. This diagram captures the full landscape of *eve* regulation, clearly delineating regions where the gene is expressed and regions where the gene is silent. It thus supports the existence of a regulatory code and accurately determines its codewords.

**11. Conclusion**

By identifying a plausible adaptor, the gene domain, our analysis supports the existence of a gene regulatory code translating specific combinations of TF concentrations into gene expression level during embryogenesis, and possibly other contexts. This viewpoint markedly differs from previously introduced gene regulatory codes that would lie in the enhancer sequence, with no adaptor involved [5-7]. The code adaptor, i.e. the gene domain forms a segregated spatial entity, in which all enhancers can at some point interact with the gene promoter, depending on surrounding TF concentrations. The encoding of gene expression achieved by the gene domain thus centrally involves genome 3D organisation. As codewords, i.e. TFs concentrations, in turn depend on the position of the nucleus in the embryo, the notion of adaptor-mediated gene regulatory code offers a novel viewpoint to investigate spatio-temporal gene patterning during development. A full understanding of this spatio-temporal gene patterning is still an open challenge attracting huge research efforts. Many mechanistic questions remain.. How does a single enhancer operate within a gene domain and how multiple enhancers cooperate ? How do TF binding sites at an enhancer allow the proper reading of TFs concentrations ? What is the expected transcriptional outcome if two enhancers were jointly active ? To this end, designing both modified cis-regulatory sequences [77] and totally synthetic enhancers [78] offer promising paths that can be supplemented with evolutionary perspectives [79].

Another challenge for the next years will be to characterise the expression phase diagram for the many genes involved in development. Drosophila is a paradigmatic example since embryos are easy to image and to genetically manipulate [80]. While microscopy allows for the determination of some gene transcription in space and time, new techniques such as single cell transcripition profiling can now be used to get a genome-wide readout [81]. These high throughput approaches have now been used in a large variety of context, inclusing human and mouse [82]. We anticipate that a combination of data analysis and modelling on these results can be used to derive quantitative information on gene expression diagrams and decipher the developmental gene regulatory code.

**Acknowledgments**

We thank Mathieu Coppey and Thomas Gregor for their critical reading of an initial version of this paper and their insightful comments.

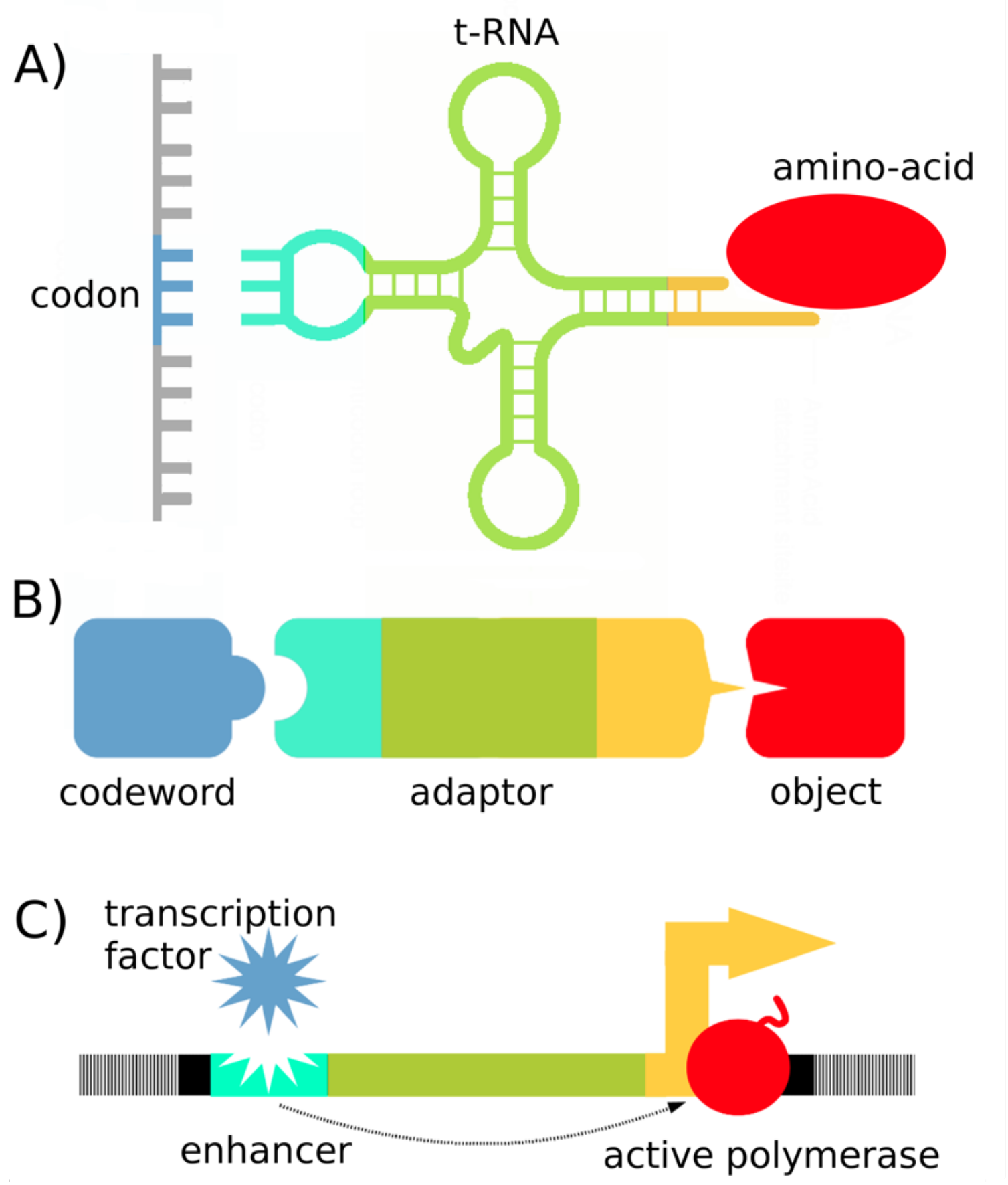

***Figure 1--- Definition of a semantic code by its adaptors.***

***A)*** *As an emblematic example of a semantic code, the genetic code is mediated by tRNAs, each bridging a codon (triplet of nucleotides, in blue) with an amino acid (red oval). Named 'adaptor' by Crick [9], a tRNA is a chimeric entity, joining a part (the anticodon, in blue-green) recognising the codon and a part (in yellow) specifically binding the amino acid.* ***B)*** *The general scheme of an adaptor emphasises the evolutionary-devised assembly, within a single molecular entity, of a site (in blue-green) specifically binding the codeword (in blue) and a distant site (in yellow) specifically binding the encoded object (in red). This adaptor defines the code.* ***C)*** *The adaptor of the gene regulatory code must comprise the distal regulatory sequences (enhancers, in blue-green) specifically binding transcription factors (in blue), and the gene promoter (in yellow) binding the active polymerase (in red).*

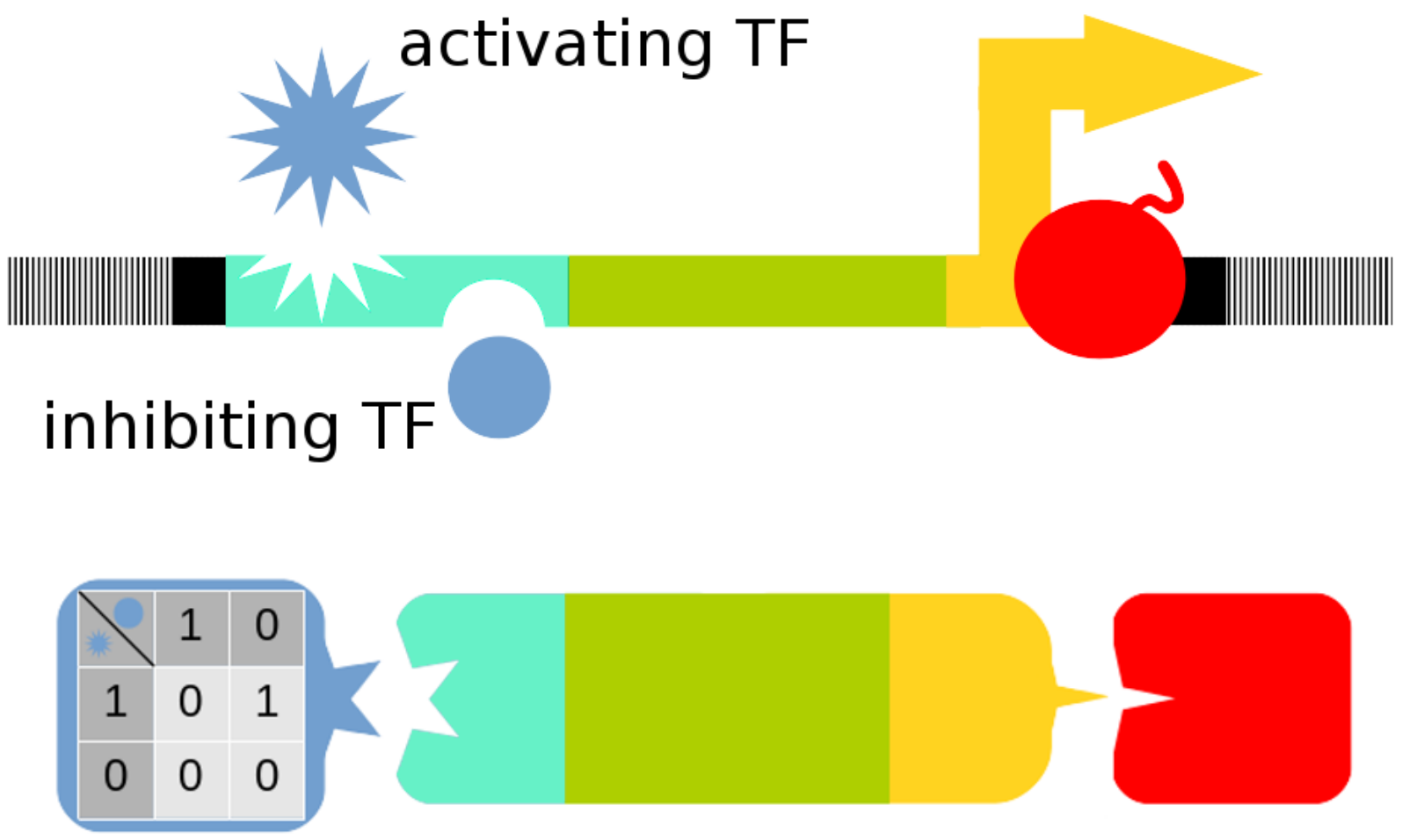

***Figure 2 --- Transcriptional control at an enhancer.***

*The transcription of a gene can be jointly activated by the binding at the enhancer of some TF (blue star) and inhibited by the binding of some other TF (blue disk). Considering an all-or-none control for simplicity, the pair of TF concentrations acts on the enhancer in a way that triggers transcription when the value given by the truth table is 1, while the gene remains silent when the value is 0. Transcription regulation at the enhancer is thus encoded in a combinatorial codeword, here (1,0). In real situations, the regulation involves TF concentration values (not only TF presence or absence as sketched here) and the transcriptional response is gradual.*

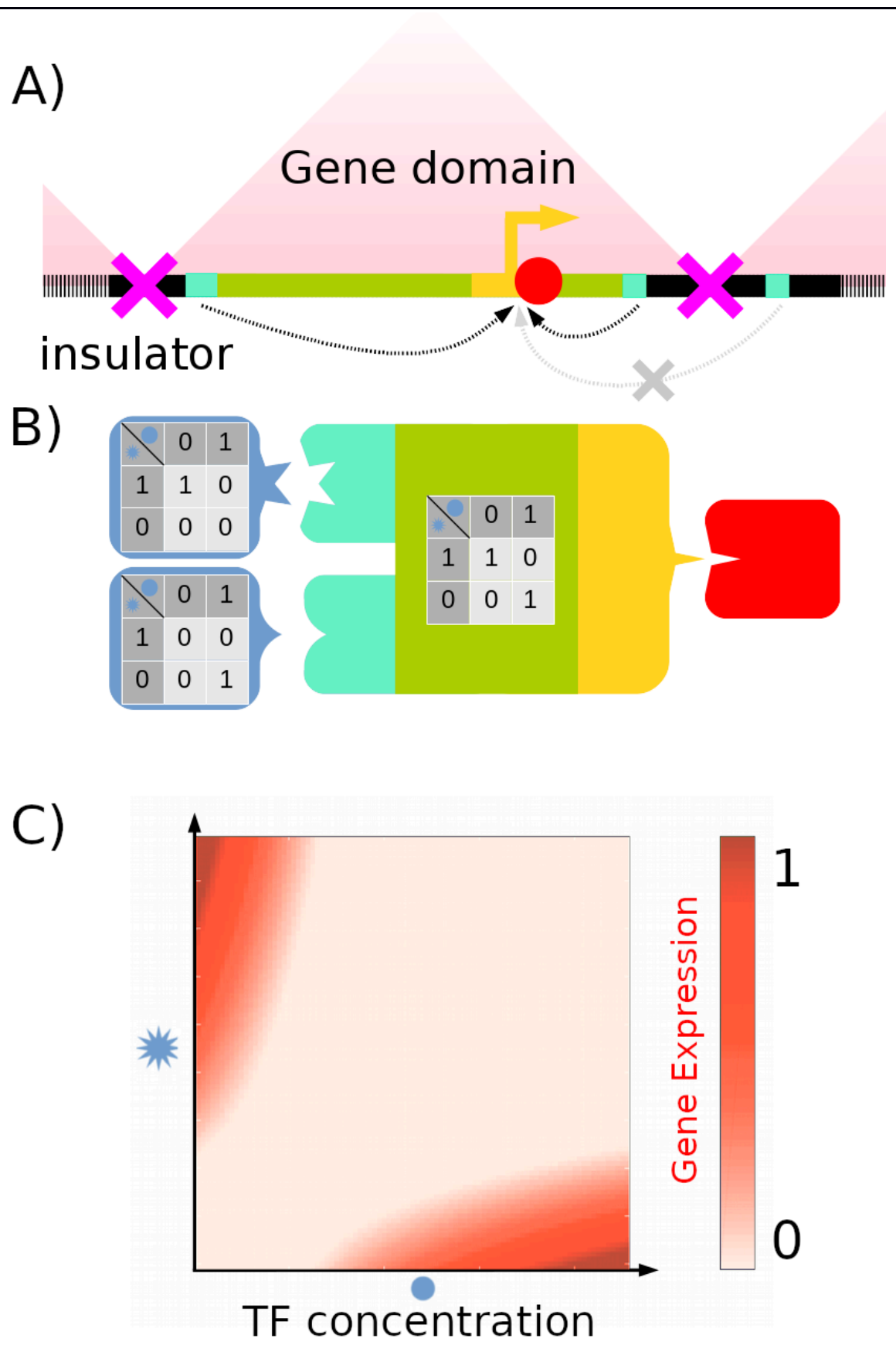

***Figure 3 -- Gene domain as a plausible adaptor of a gene regulatory code during development***

***A)*** *In general, the transcription of genes that are differentially expressed in space and time, typically developmental genes, is controlled by multiple enhancers (in blue-green) within a gene-centred genomic region delineated by insulators (magenta crosses) that we call the 'gene domain'. Contact frequencies between the different regions of the domain, determined by genome-wide chromosome conformation capture experiments (Hi-C), are depicted with a pink colour-scale on top of the domain (to be read as the diagonal of a Hi-C contact map). The analysis of these contact frequencies reveals the 3D folding of the gene domain in the nuclear space and the absence of physical contacts between genomic loci separated by insulators, preventing an action of the enhancers outside the domain.* ***B)*** *The gene domain acts as an adaptor. Each enhancer responds to TF concentrations according to a specific combinatorics, summarised (in the simplest case of an all-or-none control) in a truth table, as detailed in Fig. 2. The truth tables of the different enhancers are then integrated at the domain level, as sketched by the truth table in the middle. In this example, each TF can be either activating or inhibiting*

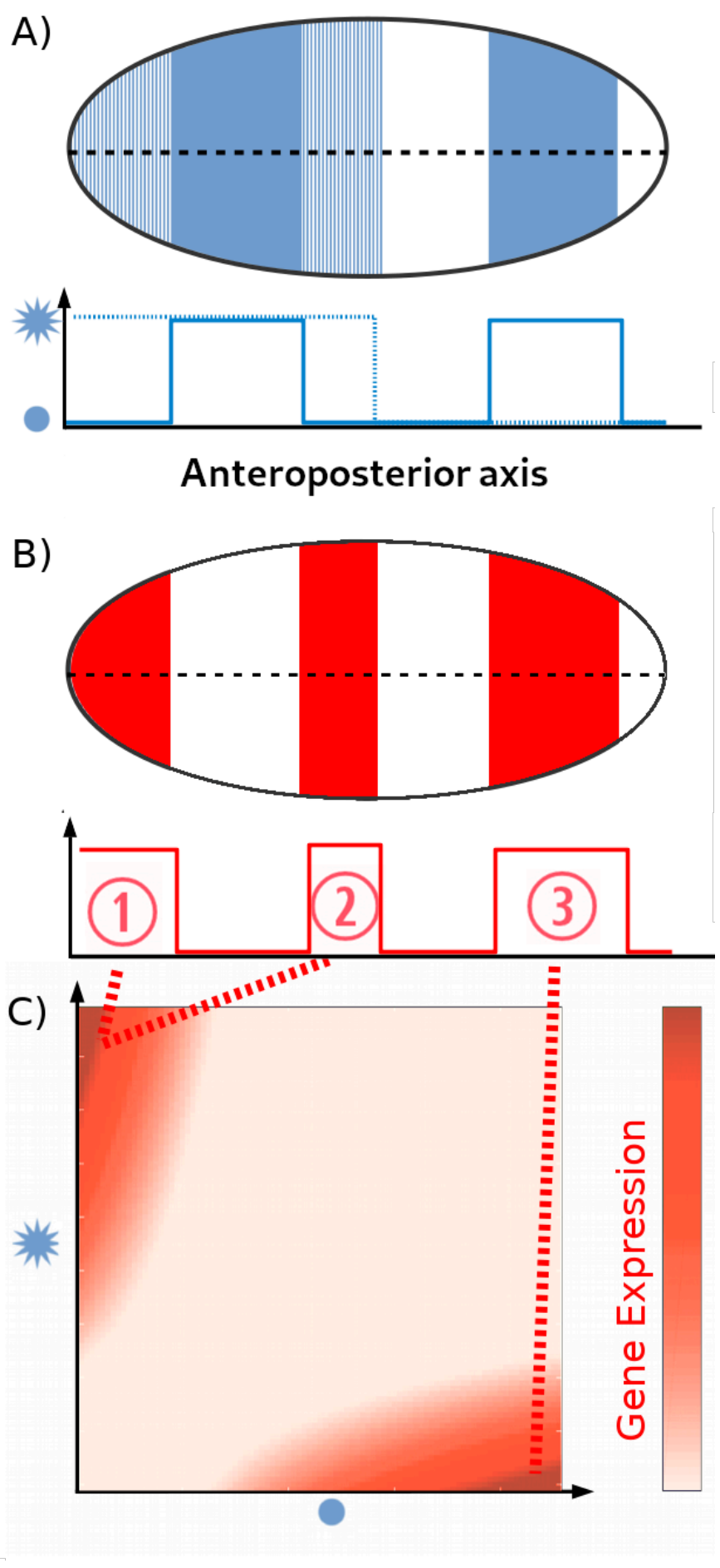

***Figure 4 --- Encoding of the early patterning of Drosophila embryo (scheme).***

***A)*** *Spatially extending the abstract logical scheme of Fig. 3, TF concentrations are now varying along the anterior-posterior axis of an embryo (oval shape). A star-shaped TF (dotted concentration curve) and a disk-shaped TF (bold concentration curve) jointly encode the local regulation of a downstream developmental gene according to the integrated truth table in Fig. 3B.* ***B)*** *Accordingly, the gene expression level (in general measured as a mRNA level) follows the variation of the two TF concentrations and takes different value in successive regions of the embryo (red stripes, and red curve along the anterior-posterior axis). As in Fig. 3, an all-or-none control has been considered for simplicity, however what occurs in reality is a gradual regulation based on continuous concentration values (see panel C)* ***C)*** *The expression phase diagram of the gene, Fig. 3C, can be drawn experimentally by scanning the various combinations of TF levels. See Fig. 6 for a similar diagram based on real data.*

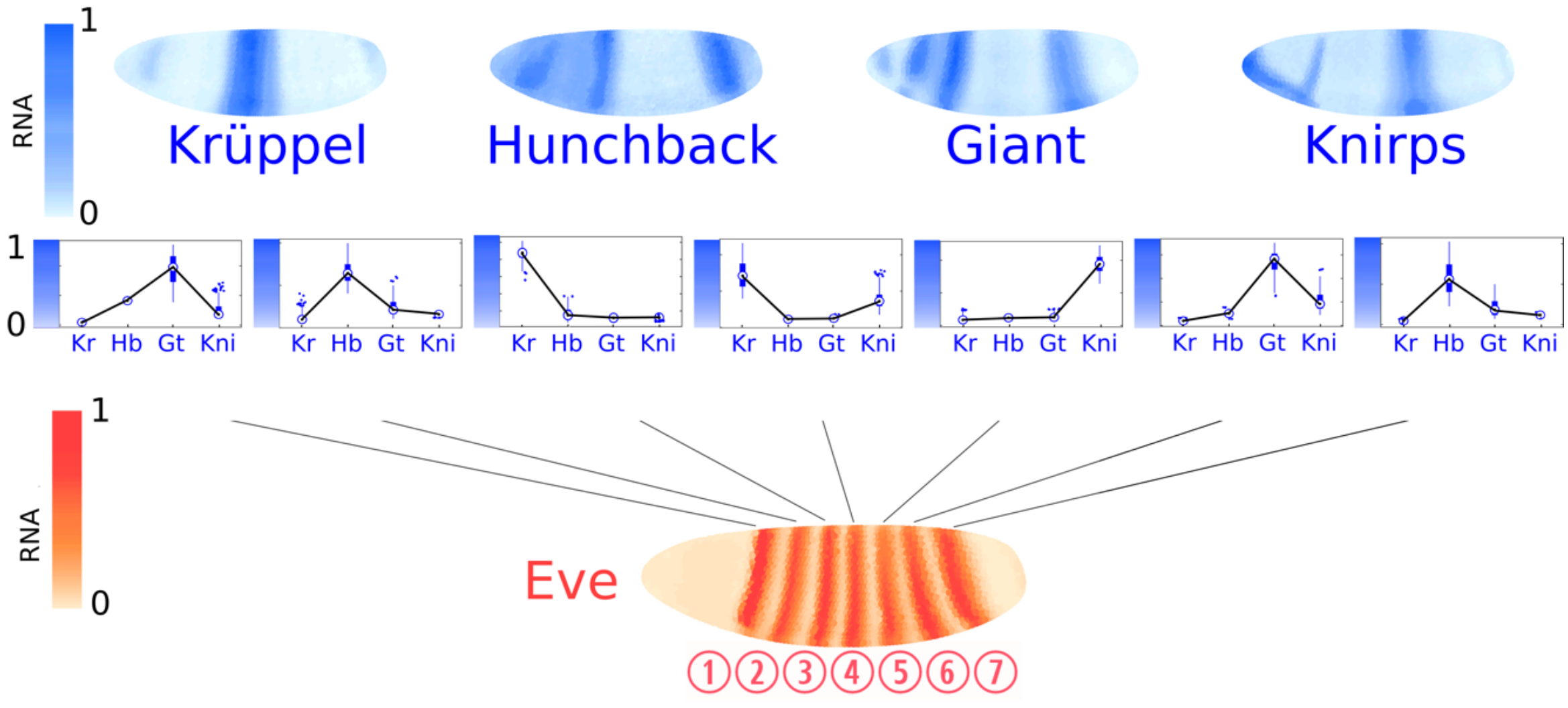

***Figure 5 --- Gene regulatory code for*** **eve** ***gene expression during Drosophila embryogenesis***

*Experimental measures of the concentrations of the four gap-gene products (Krüpple-Kr, Hunchback-Hb, Giant-Gt and Knirps-Kni, top) and the striped-pattern of* eve *gene product in Drosophila embryo (bottom). The four gap proteins act as TFs for* eve. *From these data, we quantified the distribution of the four gap-gene products in each of the seven* eve *stripe (middle). The result for each stripe is presented in the form of a boxplot (average, standard deviation and outliers). This analysis shows that each stripe is encoded by a specific combination of the four gap-gene levels, acting as a codeword in the* eve *gene regulatory code. Data from [21], normalised mRNA levels.*

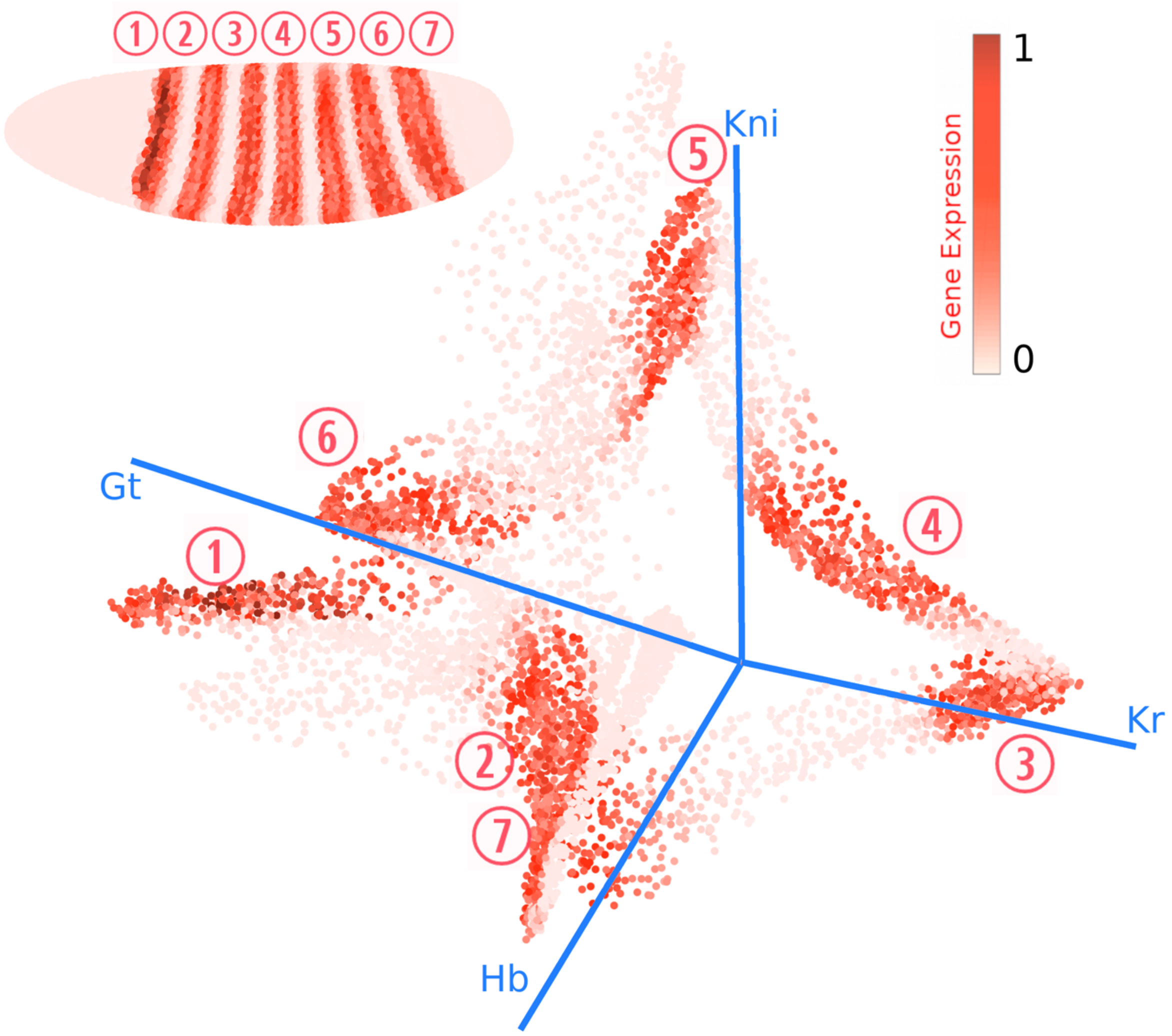

***Figure 6 --- Encoding of the early patterning of Drosophila embryo (gene*** **eve*****, real data).***

*A refined view of the regulatory code of the gene* eve *is provided by its expression phase diagram, describing as a heat map its expression level as a function of the four gap-gene mRNA levels (Kr, Hb, Gt, Kni, see Fig. 5), measured in each nucleus of the embryo. Compared to the notion sketched in Fig. 4C, the TF concentration space is now four-dimensional. To be represented on this page, the phase diagram has been projected into a 2-dimensional representation using a classical multi-dimensional scaling method. Accordingly, the four bold lines are not coordinate axes; they only correspond to codewords with only one of the four TFs being present. The spatial resolution of the measures enables an extensive scan of the TF concentration space (overall about 6000 nuclei, i.e. 6000 dots). This expression phase diagram delineates regions ('phases') in the TF concentration space where* eve *is expressed (stripes numbered from 1 to 7) and regions where this gene is inactive. Data from [21], normalised mRNA levels.*